\documentclass[conference]{IEEEtran}

\usepackage[letterpaper, left=1in, right=1in, bottom=1in, top=0.75in]{geometry}
\usepackage{comment}
\usepackage{fancyhdr}
\usepackage{cite}
\usepackage{graphicx}
\usepackage{color}
\usepackage{array}
\usepackage{epsfig}
\usepackage{amssymb}
\usepackage{amsthm}
\usepackage{multirow}
\usepackage{amsmath,mathtools}
\usepackage{algorithm}
\usepackage[noend]{algpseudocode} 
\makeatletter 
\def\BState{\State\hskip-\ALG@thistlm}

\newtheorem{Proposition}{Proposition}

\begin{document}
%
\title{Compact QC-LDPC Block and SC-LDPC Convolutional Codes for Low-Latency Communications}

%
\author{\IEEEauthorblockN{Massimo Battaglioni\IEEEauthorrefmark{1},
Alireza Tasdighi\IEEEauthorrefmark{2},
Marco Baldi\IEEEauthorrefmark{1},\\
Mohammad H. Tadayon\IEEEauthorrefmark{3} and
Franco Chiaraluce\IEEEauthorrefmark{1}}
\IEEEauthorblockA{\IEEEauthorrefmark{1}Dipartimento di Ingegneria dell'Informazione, Universit\`a Politecnica delle Marche, Ancona, Italy\\
Email: m.battaglioni@pm.univpm.it, \{m.baldi,f.chiaraluce\}@univpm.it}
\IEEEauthorblockA{\IEEEauthorrefmark{2}Department of Mathematics and Computer Science, Amirkabir University of Technology, Tehran, Iran\\
	Email: a.tasdighi@aut.ac.ir}
\IEEEauthorblockA{\IEEEauthorrefmark{3}Iran Telecommunication Research Center (ITRC),
	Tehran, Iran\\
	Email: tadayon@itrc.ac.ir}
}


\maketitle

\begin{abstract}
Low decoding latency and complexity are two important requirements of channel codes used in many applications, like machine-to-machine communications. In this paper, we show how these requirements can be fulfilled by using some special quasi-cyclic low-density parity-check block codes and spatially coupled low-density parity-check convolutional codes that we denote as \textit{compact}.
They are defined by parity-check matrices designed according to a recent approach based on sequentially multiplied columns. This method allows obtaining codes with girth up to 12. Many numerical examples of practical codes are provided.
\end{abstract}


\IEEEpeerreviewmaketitle

\section{Introduction}

Fast and reliable transmissions of short packets are a prerequisite for many modern applications, like machine-to-machine (M2M) communications \cite{Durisi2016}. 
Channel coding is commonly used for transmission reliability; however, constrained resources enforce the use of codes that can be encoded and decoded with low complexity \cite{BattaglioniPao}.
At the same time, achieving low latency is crucial in this kind of applications, due to their real-time requirements.

Quasi-cyclic low-density parity-check (QC-LDPC) block codes fit well this scenario, as they can be encoded and decoded with low-complexity, hardware-oriented techniques \cite{Li1, Lan2007}.
Iterative algorithms used for their decoding are adversely affected by the presence of short cycles in their associated Tanner graphs. Therefore, the minimum cycle length, also known as \textit{girth} (and denoted by $g$ afterwards), should be kept as large as possible.
QC-LDPC block codes are also the basis for the design of spatially coupled low-density parity-check convolutional codes (SC-LDPC-CCs). Due to their infinite length, SC-LDPC-CCs may seem unsuitable for resource-constrained contexts. However, sliding window (SW) decoding \cite{Felt1,Lentmaier2005} of SC-LDPC-CCs with short constraint length can be performed over short windows, thus resulting in very good performance, often better than that of their block code counterparts.

Motivated by these arguments, in this paper we provide design examples and assess the performance of either QC-LDPC block codes with smaller blocklength or SC-LDPC-CCs with smaller constraint length than those with comparable girth available in the literature.
These codes are designed according to the approach we have recently introduced in \cite{Tadayon2018}, and are denoted as \textit{compact codes} to encompass block and convolutional LDPC codes with these features in one word. 

The method in \cite{Tadayon2018} is based on sequentially multiplied columns (SMCs) and, to the best of our knowledge, produces the most compact QC-LDPC block codes and SC-LDPC-CCs with $g=10, 12$ currently available in the literature.
However, in \cite{Tadayon2018} a single design example is proposed.
Here we design several codes with different rates and girth, and generalize the approach to the cases of $g=6,8$.
Moreover, we relate the blocklength (for QC-LDPC block codes) and the constraint length (for SC-LDPC-CCs) of these codes to the latency and complexity of the decoding algorithms.
Many theoretical lower bounds on the blocklength (constraint length) of QC-LDPC block codes (SC-LDPC-CCs) for several values of the girth have been proposed (see, for example, \cite{Hagi2009,ATasdighi1,Karimi2013,Amirzadeh1} for QC-LDPC block codes and \cite{MBAT2017} for SC-LDPC-CCs). Compared with numerical results, these bounds are tight when $g=6,8$, but provide a loose indication when $g=10,12$. For this reason, we focus on the latter cases, quantifying the improvement achieved by the newly designed codes over previous solutions.

The remainder of the paper is organized as follows. In Section \ref{Sec2} we briefly remind the basic notions concerning QC-LDPC block codes and SC-LDPC-CCs. In Section \ref{Sec3} we recall the SMC assumption. In Section \ref{Sec4} we discuss the latency and complexity of the considered decoding algorithms. Section \ref{Sec5} provides numerical results. Finally, Section \ref{Sec6} concludes the paper.

\section{Notation}
\label{Sec2}

\subsection{CPM-based QC-LDPC block codes}

We consider a special class of QC-LDPC block codes defined through a parity-check matrix formed by $m \times n$ circulant permutation matrices (CPMs) with size $N \times N$, where $N$ is known as the \textit{lifting degree} of the code.
Each CPM is denoted as $\mathbf{I}(p_{ij})$, $0 \leq i \leq m - 1$, $0 \leq j \leq n - 1$, and is obtained by cyclically shifting all the rows of the identity matrix by $p_{ij}$ positions, with $0 \leq p_{ij} \leq N - 1$. The code length is $L=nN$. QC-LDPC block codes can be equivalently represented through their {\em exponent matrix} $\mathbf{P}$, whose entries are the integer values $p_{ij}$.

It is shown in \cite{MFossorier1} that a necessary and sufficient condition for the existence of a cycle with length $2k$ in the Tanner graph of a QC-LDPC block code is
\begin{equation}
\sum_{i=0}^{k-1} \left( p_{m_{i}n_{i}} - p_{m_{i}n_{i+1}} \right) = 0  \mod N , 
\label{fore}
\end{equation}   
where $n_{k}=n_{0}$, $m_{i} \neq m_{i+1}$, $n_{i} \neq n_{i+1}$.


Based on \eqref{fore}, let us introduce {\em avoidable} and {\em strictly avoidable} cycles for CPM-based QC-LDPC block codes. The former occur when $\sum_{i=0}^{k-1} \left( p_{m_{i}n_{i}} - p_{m_{i}n_{i+1}} \right) = \beta N$, $\beta > 0$. 
For the latter instead we have $\sum_{i=0}^{k-1} \left( p_{m_{i}n_{i}} - p_{m_{i}n_{i+1}} \right)=0$.

\subsection{SC-LDPC-CCs}

Besides QC-LDPC block codes, we consider time-invariant SC-LDPC-CCs, which are defined through a semi-infinite parity-check matrix in the form
\begin{equation}
\mathbf{H} = \left[\begin{array}{cccccc}
\arraycolsep=1.4pt\def\arraystretch{5pt}
\mathbf{H}_0 				& \mathbf{0} 							& \mathbf{0} 							& \ddots \\
\mathbf{H}_1				& \mathbf{H}_0					& \mathbf{0} 							& \ddots \\
\vdots				& \mathbf{H}_1					& \mathbf{H}_0					& \ddots \\
\mathbf{H}_{m_h}			& 	\vdots		& \mathbf{H}_1 				& \ddots \\
\mathbf{0} 						& \mathbf{H}_{m_h}	& \vdots					& \ddots \\
\mathbf{0} 						& \mathbf{0} 							& \mathbf{H}_{m_h}	& \ddots \\
\vdots				& \vdots					& \vdots					& \ddots \\
\end{array}\right],
\label{eq:Hconv}
\end{equation}
where each block $\mathbf{H}_i$, $i = 0, 1, 2, \ldots, m_h$, is a binary matrix with size $c \times a$.
The syndrome former matrix is $\mathbf{H_s} = \left[ \mathbf{H}_0^T | \mathbf{H}_1^T | \mathbf{H}_2^T | \ldots | \mathbf{H}_{m_h}^T \right]$, where $^T$ denotes transposition; its size is $a\times (m_h+1)c$. 
According to \eqref{eq:Hconv}, the code has asymptotic rate $R = \frac{a-c}{a}$.
The height of the non-zero diagonal band in \eqref{eq:Hconv} instead gives the syndrome former memory order $m_h$, and the code syndrome former constraint length is defined as $v_s = (m_h + 1) a$.

\subsection{Link between QC-LDPC block codes and SC-LDPC-CCs}

A common representation of the syndrome former matrix $\mathbf{H_s}$ of an SC-LDPC-CC has polynomials in $F_2[x]$ as its entries, where $F_2[x]$ is the ring of polynomials with coefficients in the Galois field $F_2$. In this case, the code is described by a $c \times a$ {\em symbolic matrix}
\begin{equation}
\mathbf{H}(x)=\left[\begin{array}{ccc}
h_{0,0}(x) & \ldots & h_{0,a-1}(x)\\
\vdots & \ddots &\vdots  \\
h_{c-1,0}(x) & \ldots & h_{c-1,a-1}(x)\end{array}\right],
\label{eq:Hx}
\end{equation}
where each $h_{i,j}(x)$, $i = 0, 1, 2, \ldots, c-1$, $j = 0, 1, 2, \ldots, a-1$, is a polynomial in $F_2[x]$.
The code representation based on $\mathbf{H_s}$ can be converted into that based on $\mathbf{H}(x)$ by using the following expression
\begin{equation}
h_{i,j}(x)=\sum_{m=0}^{m_h} h_{m}^{(i,j)} x^{m},
\label{eq:bintopol}
\end{equation}
where $h_{m}^{(i,j)}$ is the $(i, j)$-th entry of the matrix $\mathbf{H}_m$, the latter being the transpose of the $m$-th block of $\mathbf{H_s}$.

We focus on codes described by a symbolic parity-check matrix containing only polynomials with unitary weight, also known as \textit{monomial codes}. In this case, $\mathbf{H}(x)$ can be described through an exponent matrix in the form 
\begin{equation}
\mathbf{P}=\left[\begin{array}{llll}
p_{0,0} & p_{0,1} & \ldots & p_{0,a-1}\\
p_{1,0} & p_{1,1} & \ldots & p_{1,a-1}\\
\vdots & \vdots & \ddots & \vdots\\
p_{c-1,0} & p_{c-1,1} & \ldots & p_{c-1,a-1}\end{array}\right],
\label{eq:expomatrix}
\end{equation}
where $p_{i,j}$ is the exponent of the (only) non-null term in $h_{i,j}(x)$. The syndrome former memory order $m_h$ is the largest difference, in absolute value, between any two elements of $\mathbf{P}$. 

\section{Code design via SCM}
\label{Sec3}

In this section we recall the basic assumptions of the design method proposed in \cite{Tadayon2018}.
The design of the parity-check matrix of a QC-LDPC block code with lifting degree $N$ starts from an exponent matrix having the following form (SMC assumption)
\begin{equation}
\mathbf{P}^{\mathrm{SMC}}_{m\times n}=\left[\begin{array}{c|c|c|c|c}
\vec{0} & \vec{P}_{1} & \gamma_{2} \otimes \vec{P}_{1} & \ldots &  \gamma_{n-1} \otimes \vec{P}_{1}
\end{array}\right],
\label{eq:SMCexpomatrix}
\end{equation}
with $m,n,N\in \mathbb{N}$, $m<n\leq N$, and $\vec{0}$ and $\vec{P}_{1}$ being column vectors with $m$ entries in $ \{0,\ldots, N-1\}$. 
The vector $\vec{0}$ is filled with all zero entries, while the entries of the vector $\vec{P}_{1}$ are chosen as follows: the first entry is zero, the second entry is one and the other entries are chosen in $\{2,\ldots,N - 1\}$ in increasing order.
Then, the subsequent vectors have the form $\gamma_{j} \otimes \vec{P}_{1}$ ($j = 2,\ldots, n - 1$), where $\otimes$ denotes multiplication mod $N$, and are computed from $\vec{P}_{1}$ through sequential multiplications by the coefficients $\gamma_{j}\in \{2,\ldots ,N - 1\}$ such that $\gamma_{j}<\gamma_{j + 1}$.
The following proposition holds, which generalizes \cite[Proposition 1]{Tadayon2018}.
\begin{Proposition}
\label{SMCproposition}
Let
$\mathbf{P}^{\mathrm{SMC}}_{m\times n}$ be the exponent matrix of a QC-LDPC block code $C$ as defined in (\ref{eq:SMCexpomatrix}). Suppose that the Tanner graph associated to the submatrix $\left[\begin{array}{@{}c@{}|@{}c@{}} \vec{0} & \vec{P}_{1} \end{array}\right]$ contains no strictly avoidable cycles of length up to $\lambda$, $\lambda \in \{4,\ldots,10\}$. Then, the Tanner graph of $C$ has no strictly avoidable cycle of length up to $\lambda$ for sufficiently large $N$ and a proper choice of $\gamma_j$'s.
\end{Proposition}

\begin{IEEEproof}
Similar to the proof of \cite[Proposition 1]{Tadayon2018} and omitted here for saving space.
\end{IEEEproof}

The pseudocode for the algorithm that finds the smallest possible $\gamma_j$, $j=2,\ldots, n-1$, leading to the desired girth can be found in \cite[Algorithm 1]{Tadayon2018}.

\section{Decoding latency and complexity} \label{Sec4}

In this section we discuss the latency and complexity of the considered decoding algorithms.

\subsection{QC-LDPC block codes}

QC-LDPC block codes can be efficiently decoded by means of belief propagation (BP) algorithms. These algorithms must be executed over the whole length of the codeword, that is, $L$. So, the decoding latency, expressed as the number of bits that must be awaited before the decoding process starts, is
\begin{equation}
\Lambda_{\mathrm{BP}} = L = nN.
\label{eq:lambdaBP}
\end{equation}

The per-output-bit decoding complexity can be measured as the number of binary operations required per decoding instance per output bit. We refer to the implementation of the BP decoder proposed in \cite{Hu2001} and define the average per-output-bit decoding complexity as

\begin{equation}
\begin{aligned}
\Gamma_{\mathrm{BP}}=\frac{LI_{\mathrm{avg}}f(m,R)}{L}=I_{\mathrm{avg}}f(m,R),
\end{aligned}
\label{eq:complex}
\end{equation}
where $I_{\mathrm{avg}}$ is the average number of decoding iterations and $f(x,R)=[8(8x+12R-11)+x]$.
Notice that $\Lambda_{\mathrm{BP}}$ benefits from a reduction in the code blocklength, whereas the per-output-bit complexity does not depend on $L$. 

\subsection{SC-LDPC convolutional codes}\label{subsec:4b}

SW iterative algorithms perform BP over a window including $W$ blocks of $a$ bits each, and then let this window slide forward by $a$ bits before starting over again. For each decoding window position, the SW decoder gives the first $a$ decoded bits as output, before letting the window shift forward by $a$ bits. To ensure that the performance loss due to the non-infinite size of the sliding window is negligible, the number of blocks has to be $W=\alpha(m_h+1)$, with $\alpha \geq 5$. By using this value of $W$, we can express the decoding latency ($\Lambda_{\mathrm{SW}}$) and average per-output-bit complexity ($\Gamma_{\mathrm{SW}}$) of a SW decoder as
\begin{equation}
\begin{cases}
\Lambda_{\mathrm{SW}} = Wa = \alpha(m_h+1)a,\\
\begin{aligned}
\Gamma_{\mathrm{SW}} &= \frac{WaI_{\mathrm{avg}} f(c,R)}{a}=\\
&=\alpha(m_h+1)I_{\mathrm{avg}} f(c,R).
\end{aligned}
\end{cases}
\label{cas}
\end{equation}

Note that SC-LDPC-CCs characterized by small values of $m_h$ can be decoded with small window sizes. According to \eqref{cas}, this results in a reduction of both the decoding latency and per-output-bit complexity.

\section{Numerical results} \label{Sec5}

\begin{table}[t]
\setlength{\tabcolsep}{.7 pt}
\centering
\caption{Exponent Matrices of the Shortest QC-LDPC Block Codes with Girth 10, Constructed from a $3\times n$ Fully-Connected Base Graph}
\label{Table:TTab1}
\vspace{-.7em}
\begingroup\fontsize{6.5pt}{7pt}
\begin{tabular}{|@{}c@{}|@{}c@{}|@{}c@{}|@{}l@{}|}
\hline
  $n$ & $R$ & $N$ & Exponent Matrix   \\
  \hline \hline
  $4$ & $0.263$ &  $\renewcommand{\arraystretch}{0.3}\begin{array}{@{}c@{}}
   37\\
   (37\text{\cite{Bocharova1}} ,\text{\cite{ATasdighi1}} ) \end{array}$ & $\renewcommand{\arraystretch}{0.3}\begin{array}{l}
1, 3, 24\\
27, 7, 19
  \end{array} $ \\
\hline
  $5$ & $0.406$ &  $\renewcommand{\arraystretch}{0.3}\begin{array}{@{}c@{}}
   61\\
   (61\text{\cite{Bocharova1}} ,\text{\cite{ATasdighi1}} ) \end{array}$ & $\renewcommand{\arraystretch}{0.3}\begin{array}{l}
1, 3, 21, 55\\
5, 15, 44, 31
  \end{array}$ \\
\hline
  $6$ & $0.503$ &  $\renewcommand{\arraystretch}{0.3}\begin{array}{@{}c@{}}
   91\\
   (91 \text{\cite{ATasdighi1}} ) \end{array}$ & $\renewcommand{\arraystretch}{0.3}\begin{array}{l}
1, 3, 7, 25, 38\\
17, 51, 28, 61, 9
  \end{array}$ \\
\hline
  $7$ & $0.573$ &  $\renewcommand{\arraystretch}{0.3}\begin{array}{@{}c@{}}
   139\\
   (145 \text{\cite{Amirzadeh1}} ) \end{array}$ & $\renewcommand{\arraystretch}{0.3}\begin{array}{l}
1, 3, 8, 25, 34, 95 \\
97, 13, 81, 62, 101, 41 
  \end{array}$ \\
\hline
  $8$ & $0.626$ &  $\renewcommand{\arraystretch}{0.3}\begin{array}{@{}c@{}}
   181\\
   (211 \text{\cite{Amirzadeh1}} ) \end{array}$ & $\renewcommand{\arraystretch}{0.3}\begin{array}{l}
1, 3, 69, 120, 129, 141, 156 \\
133, 37, 127, 32, 143, 110, 114
  \end{array}$ \\
\hline
  $9$ & $0.667$ &  $\renewcommand{\arraystretch}{0.3}\begin{array}{@{}c@{}}
   241\\
   (319 \text{\cite{Bocharova1}} ) \end{array}$ & $\renewcommand{\arraystretch}{0.3}\begin{array}{l}
1, 3, 13, 88, 114, 182, 217, 223 \\
16, 48, 208, 203, 137, 20, 98, 194
  \end{array}$ \\
\hline
  $10$ & $0.700$ &  $\renewcommand{\arraystretch}{0.3}\begin{array}{@{}c@{}}
   313\\
   (430 \text{\cite{Bocharova1}} ) \end{array}$ & $\renewcommand{\arraystretch}{0.3}\begin{array}{l}
1, 3, 7, 15, 49, 66, 189, 220, 292 \\
215, 19, 253, 95, 206, 105, 258, 37, 180
  \end{array}$ \\
\hline
  $11$ & $0.727$ &  $\renewcommand{\arraystretch}{0.3}\begin{array}{@{}c@{}}
   397\\
   (560 \text{\cite{Bocharova1}} ) \end{array}$ & $\renewcommand{\arraystretch}{0.3}\begin{array}{l}
1, 3, 7, 15, 62, 127, 146, 183, 209, 301 \\
35, 105, 245, 128, 185, 78, 346, 53, 169, 213
  \end{array}$ \\
  
\hline
  $12$ & $0.750$ &  $\renewcommand{\arraystretch}{0.3}\begin{array}{@{}c@{}}
   523\\
   (737 \text{\cite{Bocharova1}} ) \end{array}$ & $\renewcommand{\arraystretch}{0.3}\begin{array}{l}
1, 3, 7, 12, 25, 58, 288, 320, 392, 429, 437 \\
463, 343, 103, 326, 69, 181, 502, 151, 15,...\\ 410, 453
  \end{array}$ \\
\hline
\end{tabular}
\endgroup
\vspace{-.7em}
\end{table}

\begin{table}[t]
\setlength{\tabcolsep}{.7 pt}
\centering
\caption{Exponent Matrices of the Shortest QC-LDPC Block Codes with Girth 10, Constructed from a $4\times n$ Fully-Connected Base Graph}
\label{Table:TTab2}
\vspace{-.7em}
\begingroup\fontsize{6.5pt}{7pt}
\begin{tabular}{|@{}c@{}|@{}c@{}|@{}c@{}|@{}l@{}|}
\hline
  $n$ & $R$ & $N$ & Exponent Matrix   \\
  \hline \hline
  $5$ & $0.204$ &  $\renewcommand{\arraystretch}{0.3}\begin{array}{@{}c@{}}
   139\\
   (223\text{\cite{sullivan1}} ) \end{array}$ & $\renewcommand{\arraystretch}{0.3}\begin{array}{l}
1, 3, 30, 105\\
43, 129, 39, 67\\
61, 44, 23, 11
  \end{array} $ \\
\hline
$6$ & $0.335$ &  $\renewcommand{\arraystretch}{0.3}\begin{array}{@{}c@{}}
   241\\
   (383\text{\cite{sullivan1}} ) \end{array}$ & $\renewcommand{\arraystretch}{0.3}\begin{array}{l}
1, 3, 7, 80, 147\\
16, 48, 112, 75, 183\\
86, 17, 120, 132, 110
\end{array}$ \\
\hline
  $7$ & $0.429$ &  $\renewcommand{\arraystretch}{0.3}\begin{array}{@{}c@{}}
   307\\
   (601 \text{\cite{sullivan1}} ) \end{array}$ & $\renewcommand{\arraystretch}{0.3}\begin{array}{l}
1, 3, 12, 124, 130, 235\\
18, 54, 216, 83, 191, 239\\
211, 19, 76, 69, 107, 158
  \end{array}$ \\
\hline
  $8$ & $0.500$ &  $\renewcommand{\arraystretch}{0.3}\begin{array}{@{}c@{}}
   409\\
   (827 \text{\cite{sullivan1}} ) \end{array}$ & $\renewcommand{\arraystretch}{0.3}\begin{array}{l}
1, 3, 14, 59, 144, 180, 184\\
54, 162, 347, 323, 5, 313, 120 \\
291, 55, 393, 400, 186, 28, 374  
  \end{array}$ \\
\hline
  $9$ & $0.556$ &  $\renewcommand{\arraystretch}{0.3}\begin{array}{@{}c@{}}
   577\\
   (1223 \text{\cite{sullivan1}} ) \end{array}$ & $\renewcommand{\arraystretch}{0.3}\begin{array}{l}
1, 3, 7, 61, 85, 168, 235, 550 \\
214, 65, 344, 360, 303, 178, 91, 569\\
264, 215, 117, 525, 514, 500, 301, 373
  \end{array}$ \\
\hline
  $10$  & $0.600$ &  $\renewcommand{\arraystretch}{0.3}\begin{array}{@{}c@{}}
   787\\
   (1667 \text{\cite{sullivan1}} ) \end{array}$ & $\renewcommand{\arraystretch}{0.3}\begin{array}{l}
1, 3, 7, 20, 104, 215, 245, 702, 751 \\
380, 353, 299, 517, 170, 639, 234, 754, 486\\
127, 381, 102, 179, 616, 547, 422, 223, 150
  \end{array}$ \\
\hline
$11$  & $0.636$ &  $\renewcommand{\arraystretch}{0.3}\begin{array}{@{}c@{}}
   1039\\
   (2207 \text{\cite{sullivan1}} ) \end{array}$ & $\renewcommand{\arraystretch}{0.3}\begin{array}{l}
1, 3, 7, 12, 115, 170, 318, 388, 510, 1003 \\
141, 423, 987, 653, 630, 73, 161, 680, 219, 119\\
740, 142, 1024, 568, 941, 81, 506, 356, 243, 374
  \end{array}$ \\
\hline
$12$ & $0.666$ &  $\renewcommand{\arraystretch}{0.3}\begin{array}{@{}c@{}}
   1381\\
   (2903 \text{\cite{sullivan1}} ) \end{array}$ & $\renewcommand{\arraystretch}{0.3}\begin{array}{l}
1, 3, 7, 12, 20, 111, 716, 862, 919, 963, 1211\\
355, 1065, 1104, 117, 195, 737, 76, 809, 329,...\\ 758, 414\\
579, 356, 1291, 43, 532, 743, 264, 557, 416,...\\ 1034, 1002
  \end{array}$ \\
\hline
\end{tabular}
\endgroup
\vspace{-.7em}
\end{table}

By applying the method proposed in \cite{Tadayon2018}, we have designed several codes with girth $g=10, 12$.
The values of $N$ and $m_h$ obtained for these codes are often significantly smaller than those of other codes with the same rate and girth reported in the literature.
In particular,  we have considered $m = 3, 4$ and $n = 4, \ldots, 12$ for the QC-LDPC block codes, and $c = 3, 4$ and $a = 4, \ldots, 12$ for the SC-LDPC-CCs.
 This choice derives from the fact that codes with $m=1,2$ ($c=1,2$) entail undesirable properties which yield a very poor performance, whereas codes with $m>4$ ($c>4$) usually exhibit degraded waterfall performance and yield large decoding complexity.
We have compared the obtained values of $N$ and $m_h$ with those available in the literature. To the best of our knowledge, the design approaches that have produced till now the codes with minimum values of $N$ and $m_h$ are those reported in \cite{sullivan1,MBAT2017,Bocharova1,ATasdighi1,Amirzadeh1}.
 

The exponent matrices of the newly designed codes are reported in Tables \ref{Table:TTab1} to \ref{Table:TTab8}. The lifting degree (syndrome former memory order) of the most compact existing codes is given between square brackets. The first row and column of any QC-LDPC block code exponent matrix are filled with all-zero entries and omitted.

\begin{table}[t]
\setlength{\tabcolsep}{.7 pt}
\centering
\caption{Exponent Matrices of the Shortest QC-LDPC Block Codes with Girth 12, Constructed from a $3\times n$ Fully-Connected Base Graph}
\label{Table:TTab3}
\vspace{-.7em}
\begingroup\fontsize{6.5pt}{7pt}
\begin{tabular}{|@{}c@{}|@{}c@{}|@{}c@{}|@{}l@{}|}
\hline
  $n$ & $R$ & $N$ &Exponent Matrix   \\
  \hline \hline
  $4$  & $0.256$ &  $\renewcommand{\arraystretch}{0.3}\begin{array}{@{}c@{}}
   73\\
   (73\text{\cite{Bocharova1}} ,\text{\cite{ATasdighi1}} ) \end{array}$ & $\renewcommand{\arraystretch}{0.3}\begin{array}{l}
1, 3, 13\\
9, 27, 44
  \end{array} $ \\
\hline
$5$  & $0.402$ &  $\renewcommand{\arraystretch}{0.3}\begin{array}{@{}c@{}}
   151\\
   (156\text{\cite{ATasdighi1}} ) \end{array}$ & $\renewcommand{\arraystretch}{0.3}\begin{array}{l}
1, 3, 108, 139 \\
119, 55, 17, 82 
\end{array}$ \\
\hline
$6$ & $0.501$ &  $\renewcommand{\arraystretch}{0.3}\begin{array}{@{}c@{}}
   271\\
   (306\text{\cite{Bocharova1}} ) \end{array}$ & $\renewcommand{\arraystretch}{0.3}\begin{array}{l}
1, 3, 7, 67, 144 \\
29, 87, 203, 46, 111 
\end{array}$ \\
\hline
  $7$ & $0.572$ &  $\renewcommand{\arraystretch}{0.3}\begin{array}{@{}c@{}}
   457\\
   (566\text{\cite{Bocharova1}} ) \end{array}$ & $\renewcommand{\arraystretch}{0.3}\begin{array}{l}
1, 3, 10, 53, 311, 362 \\
134, 402, 426, 247, 87, 66 
  \end{array}$ \\
\hline
  $8$  & $0.625$ &  $\renewcommand{\arraystretch}{0.3}\begin{array}{@{}c@{}}
   691\\
   (848\text{\cite{Bocharova1}} ) \end{array}$ & $\renewcommand{\arraystretch}{0.3}\begin{array}{l}
1, 3, 9, 76, 236, 310, 539 \\
254, 71, 213, 647, 518, 657, 88 
  \end{array}$ \\
\hline
  $9$  & $0.666$ &  $\renewcommand{\arraystretch}{0.3}\begin{array}{@{}c@{}}
   991\\
   (1376\text{\cite{Bocharova1}} ) \end{array}$ & $\renewcommand{\arraystretch}{0.3}\begin{array}{l}
1, 3, 7, 134, 420, 557, 660, 672 \\
114, 342, 798, 411, 312, 74, 915, 301
  \end{array}$ \\
\hline
  $10$  & $0.700 $ &  $\renewcommand{\arraystretch}{0.3}\begin{array}{@{}c@{}}
   1447\\
   (2103\text{\cite{Bocharova1}} ) \end{array}$ & $\renewcommand{\arraystretch}{0.3}\begin{array}{l}
1, 3, 7, 22, 48, 226, 256, 489, 1190 \\
705, 668, 594, 1040, 559, 160, 1052, 359,...\\ 1137 
  \end{array}$ \\
\hline
  $11$  & $0.727 $ &  $\renewcommand{\arraystretch}{0.3}\begin{array}{@{}c@{}}
   2161\\
   (3137\text{\cite{Bocharova1}} ) \end{array}$ & $\renewcommand{\arraystretch}{0.3}\begin{array}{l}
1,3,7,12,20,313,609,700,1487,1853\\
594,1782,1997,645,1075,76,859,888,...\\1590,733
  \end{array}$ \\
\hline
\end{tabular}
\endgroup
\vspace{-.7em}
\end{table}

\begin{table}[t]
\setlength{\tabcolsep}{.7 pt}
\centering
\caption{Exponent Matrices of the Shortest QC-LDPC Block Codes with Girth 12, Constructed from a $4\times n$ Fully-Connected Base Graph}
\label{Table:TTab4}
\vspace{-.7em}
\begingroup\fontsize{6.5pt}{7pt}
\begin{tabular}{|@{}c@{}|@{}c@{}|@{}c@{}|@{}l@{}|}
\hline
  $n$ &  $R$ & $N$ &Exponent Matrix   \\
  \hline \hline
$5$ & $0.200$ &  $\renewcommand{\arraystretch}{0.3}\begin{array}{@{}c@{}}
607 \\
   (1093 \text{\cite{sullivan1}} ) \end{array}$ & $\renewcommand{\arraystretch}{0.3}\begin{array}{l}
1,4,225,536\\
211,237,129,194\\
273,485,118,41
\end{array}$ \\
\hline
$6$ & $0.333$ &  $\renewcommand{\arraystretch}{0.3}\begin{array}{@{}c@{}}
1201 \\
   (2251 \text{\cite{sullivan1}} ) \end{array}$ & $\renewcommand{\arraystretch}{0.3}\begin{array}{l}
1,4,468,470,784\\
571,1083,606,547,892\\
591,1163,358,339,959
\end{array}$ \\
\hline
  $7$ & $0.428$ &  $\renewcommand{\arraystretch}{0.3}\begin{array}{@{}c@{}}
2371\\
   (4019 \text{\cite{sullivan1}} ) \end{array}$ & $\renewcommand{\arraystretch}{0.3}\begin{array}{l}
1,4,9,655,872,2233\\
465,1860,1814,1087,39,2218\\
1736,2202,1398,1371,1094,2274
  \end{array}$ \\
\hline
\end{tabular}
\endgroup
\vspace{-.7em}
\end{table}

\begin{table}[!t]
\setlength{\tabcolsep}{.7 pt}
\centering
\caption{Lowest exponent matrices of codes with $g=10$ and $c=3$}
\label{Table:TTab5}
\vspace{-.7em}
\begingroup\fontsize{6.5pt}{7pt}
\begin{tabular}{|@{}c@{}|@{}c@{}|@{}l@{}|}
\hline
  $a$ & $m_{h}$ &~~~~~~~~~~~~~~~Exponent Matrix   \\
  \hline
  $4$ &$\renewcommand{\arraystretch}{0.3}\begin{array}{@{}c@{}}
   11\\
   (10\text{\cite{MBAT2017}} ) \end{array}$ & $\renewcommand{\arraystretch}{0.3}\begin{array}{l}
6, 11, 0, 9\\
11, 2, 0, 11\\
4, 1, 11, 0 
  \end{array} $ \\
\hline
  $5$ & $\renewcommand{\arraystretch}{0.3}\begin{array}{@{}c@{}}
   19\\
   (19\text{\cite{MBAT2017}} ) \end{array}$  & $\renewcommand{\arraystretch}{0.3}\begin{array}{l}
19, 19, 0, 17, 0\\
0, 9, 17, 6, 11\\
0, 5, 1, 16, 18
  \end{array}$ \\
\hline
  $6$ & $\renewcommand{\arraystretch}{0.3}\begin{array}{@{}c@{}}
   31\\
   (31\text{\cite{MBAT2017}} ) \end{array}$ & $\renewcommand{\arraystretch}{0.3}\begin{array}{l}
31, 31, 0, 0, 0, 30\\
0, 20, 29, 18, 14, 31\\
24, 0, 12, 7, 30, 21 
  \end{array}$ \\
\hline
$7$ & $\renewcommand{\arraystretch}{0.3}\begin{array}{@{}c@{}}
   44\\
   (53\text{\cite{MBAT2017}} ) \end{array}$ & $\renewcommand{\arraystretch}{0.3}\begin{array}{l}
0,25,44,44,25,0,12\\
44,44,13,27,0,28,44\\
40,3,37,5,44,17,0
\end{array}$ \\
\hline
$8$ & $\renewcommand{\arraystretch}{0.3}\begin{array}{@{}c@{}}
   66\\
   (76\text{\cite{MBAT2017}} ) \end{array}$ & $\renewcommand{\arraystretch}{0.3}\begin{array}{l}
0,9,12,66,65,16,66,61\\
66,0,34,25,0,0,55,11\\
48,37,0,1,66,18,9,66
\end{array}$ \\
\hline
$9$ & $\renewcommand{\arraystretch}{0.3}\begin{array}{@{}c@{}}
   88\\
   (127\text{\cite{MBAT2017}} ) \end{array}$ & $\renewcommand{\arraystretch}{0.3}\begin{array}{l}
0,0,88,0,88,76,78,0,88\\
29,88,53,73,7,83,0,59,19\\
35,15,63,16,50,0,88,33,1
\end{array}$ \\
\hline
$10$ & $\renewcommand{\arraystretch}{0.3}\begin{array}{@{}c@{}}
   124\\
   (222\text{\cite{MBAT2017}} ) \end{array}$ & $\renewcommand{\arraystretch}{0.3}\begin{array}{l}
0,0,87,62,100,100,78,53,74,67\\
46,120,42,0,4,16,0,0,124,124\\
72,19,0,76,3,79,95,124,67,0  
\end{array}$ \\
\hline
 $11$ & $\renewcommand{\arraystretch}{0.3}\begin{array}{@{}c@{}}
   177\\
   (307\text{\cite{MBAT2017}} ) \end{array}$ & $\renewcommand{\arraystretch}{0.3}\begin{array}{l}
0,41,0,100,37,38,100,0,75,22,39\\
87,158,177,0,177,0,27,100,94,27,25\\
100,0,74,7,7,130,158,158,177,31,177
\end{array}$ \\
  
\hline
  $12$ & $\renewcommand{\arraystretch}{0.3}\begin{array}{@{}c@{}}
   277\\
   (388\text{\cite{MBAT2017}} ) \end{array}$ & $\renewcommand{\arraystretch}{0.3}\begin{array}{l}
0,0,0,61,0,100,30,0,0,32,83,0\\
0,66,198,0,269,181,197,180,200,277,155,77\\
53,277,202,113,126,0,0,236,82,29,0,140
  \end{array}$ \\
\hline
\end{tabular}
\endgroup
\vspace{-.7em}
\end{table}

\begin{table}[!t]
\setlength{\tabcolsep}{.7 pt}
\centering
\caption{Lowest exponent matrices of codes with $g=10$ and $c=4$}
\label{Table:TTab6}
\vspace{-.7em}
\begingroup\fontsize{6.5pt}{7pt}
\begin{tabular}{|@{}c@{}|@{}c@{}|@{}l@{}|}
\hline
  $a$ & $m_{h}$ &~~~~~~~~~~~~~~~~~~Exponent Matrix   \\
  \hline
  $5$ & $\renewcommand{\arraystretch}{0.3}\begin{array}{@{}c@{}}
   50\\
(223 \text{\cite{sullivan1}} )  \end{array}$  & $\renewcommand{\arraystretch}{0.3}\begin{array}{l}
50,11,0,22,50\\
25,48,22,50,2\\
23,9,48,50,7\\
13,3,50,21,0
  \end{array}$ \\
\hline
  $6$ & $\renewcommand{\arraystretch}{0.3}\begin{array}{@{}c@{}}
   90\\
(383 \text{\cite{sullivan1}} ) \end{array}$ & $\renewcommand{\arraystretch}{0.3}\begin{array}{l}
0,9,0,31,90,18\\
0,39,90,0,80,90\\
73,80,67,90,3,38\\
62,0,90,78,15,6
  \end{array}$ \\
\hline
$7$ & $\renewcommand{\arraystretch}{0.3}\begin{array}{@{}c@{}}
 142\\
( 601\text{\cite{sullivan1}} ) \end{array}$ & $\renewcommand{\arraystretch}{0.3}\begin{array}{l}
0,35,13,100,75,93,100\\
96,142,142,21,0,84,18\\
75,1,68,95,142,120,41\\
100,0,15,115,13,142,96
\end{array}$ \\
\hline
$8$ & $\renewcommand{\arraystretch}{0.3}\begin{array}{@{}c@{}}
 192\\
(827 \text{\cite{sullivan1}} ) \end{array}$ & $\renewcommand{\arraystretch}{0.3}\begin{array}{l}
74,149,70,39,192,23,191,0\\
0,0,180,142,192,192,114,32\\
76,191,192,192,100,59,31,0\\
95,22,56,33,70,0,157,192
\end{array}$ \\
\hline
$9$ & $\renewcommand{\arraystretch}{0.3}\begin{array}{@{}c@{}}
 319\\
(1223 \text{\cite{sullivan1}} ) \end{array}$ & $\renewcommand{\arraystretch}{0.3}\begin{array}{l}
143,38,100,291,291,284,268,0,100\\
120,14,74,261,207,176,77,319,104\\
319,0,211,123,107,157,266,85,284\\
0,208,319,31,200,204,202,133,161
\end{array}$ \\
\hline
$10$ & $\renewcommand{\arraystretch}{0.3}\begin{array}{@{}c@{}}
   507\\
(1667 \text{\cite{sullivan1}} ) \end{array}$ & $\renewcommand{\arraystretch}{0.3}\begin{array}{l}
100,112,156,9,185,355,173,100,482,1\\
0,25,95,0,345,33,507,37,64,220\\
277,507,200,138,0,381,0,171,230,200\\
 59,148,346,507,110,452,160,36,192,336
\end{array}$ \\
\hline
 $11$ & $\renewcommand{\arraystretch}{0.3}\begin{array}{@{}c@{}}
   706\\

(2207 \text{\cite{sullivan1}} ) \end{array}$ & $\renewcommand{\arraystretch}{0.3}\begin{array}{l}
0,0,0,371,0,135,0,0,0,418,92\\ 
0,118,354,158,377,198,319,120,68,336,0\\
0,14,42,469,168,706,302,296,237,285,627\\
0,44,132,679,528,0,207,485,448,0,586
\end{array}$ \\
  
\hline
  $12$ & $\renewcommand{\arraystretch}{0.3}\begin{array}{@{}c@{}}
   1027\\
(2903 \text{\cite{sullivan1}} ) \end{array}$ & $\renewcommand{\arraystretch}{0.3}\begin{array}{l}
88,100,0,0,0,215,930,0,0,0,0,142\\
100,584,47,554,152,0,856,1000,862,146,199,12\\
1018,108,926,0,914,658,331,896,241,165,1027,378\\
138,0,981,381,1012,27,902,368,564,300,605,834
\end{array}$ \\
\hline
\end{tabular}
\endgroup
\vspace{-.7em}
\end{table}

\begin{table}[!t]
\setlength{\tabcolsep}{.7 pt}
\centering
\caption{Lowest exponent matrices of codes with $g=12$ and $c=3$}
\label{Table:TTab7}
\vspace{-.7em}
\begingroup\fontsize{6.5pt}{7pt}
\begin{tabular}{|c|c|l|}
\hline
  $a$ & $m_{h}$ &~~~~~~~~~~~~Exponent Matrix   \\
  \hline
  $4$ & $\renewcommand{\arraystretch}{0.3}\begin{array}{@{}c@{}}
   20\\
   (19\text{\cite{MBAT2017}} ) \end{array}$ & $\renewcommand{\arraystretch}{0.3}\begin{array}{l}
4, 18, 0, 20\\
0, 20, 20, 12\\
20, 15, 13, 0
  \end{array} $ \\
\hline
  $5$ & $\renewcommand{\arraystretch}{0.3}\begin{array}{@{}c@{}}
   40\\
   (42\text{\cite{MBAT2017}} ) \end{array}$  & $\renewcommand{\arraystretch}{0.3}\begin{array}{l}
0,0,0,40,40\\
25,30,40,1,5\\
40,31,13,14,37
  \end{array}$ \\
\hline
  $6$ & $\renewcommand{\arraystretch}{0.3}\begin{array}{@{}c@{}}
   86\\
   (108\text{\cite{MBAT2017}} ) \end{array}$ & $\renewcommand{\arraystretch}{0.3}\begin{array}{l}
0,82,80,63,58,55\\
46,86,0,86,0,15\\
62,10,11,0,85,62
  \end{array}$ \\
\hline
$7$ & $\renewcommand{\arraystretch}{0.3}\begin{array}{@{}c@{}}
   165\\
   (220\text{\cite{MBAT2017}} ) \end{array}$ & $\renewcommand{\arraystretch}{0.3}\begin{array}{l}
148,100,48,165,11,0,160\\
0,0,44,40,122,156,22\\
163,149,165,63,0,78,144
\end{array}$ \\
\hline
$8$ & $\renewcommand{\arraystretch}{0.3}\begin{array}{@{}c@{}}
   297\\
   (442\text{\cite{MBAT2017}} ) \end{array}$ & $\renewcommand{\arraystretch}{0.3}\begin{array}{l}
35,76,60,297,297,0,6,135\\
98,99,3,0,84,297,107,59\\
51,297,0,85,0,26,0,86
\end{array}$ \\
\hline
$9$ & $\renewcommand{\arraystretch}{0.3}\begin{array}{@{}c@{}}
   468\\
   (852\text{\cite{MBAT2017}} ) \end{array}$ & $\renewcommand{\arraystretch}{0.3}\begin{array}{l}
0,0,0,50,0,0,316,104,388\\
0,156,468,151,93,114,0,0,174\\
328,274,166,0,29,441,296,468,104
\end{array}$ \\
\hline
$10$ & $\renewcommand{\arraystretch}{0.3}\begin{array}{@{}c@{}}
   797\\
   (1231\text{\cite{MBAT2017}} ) \end{array}$ & $\renewcommand{\arraystretch}{0.3}\begin{array}{l}
0,0,0,0,0,10,0,0,18,56\\
0,663,542,300,116,0,797,429,97,411\\
0,34,102,238,748,195,449,22,727,0
\end{array}$ \\
\hline
$11$ & $\renewcommand{\arraystretch}{0.3}\begin{array}{@{}c@{}}
1075\\
(1958\text{\cite{Bocharova1}} ) \end{array}$ & $\renewcommand{\arraystretch}{0.3}\begin{array}{l}
0,0,379,164,0,0,0,0,0,674,308\\
0,594,0,0,645,1075,76,859,888,103,1041\\
0,1,382,171,12,20,313,609,700,0,0
\end{array}$ \\
\hline
\end{tabular}
\endgroup
\vspace{-.7em}
\end{table}

\begin{table}[!t]
\setlength{\tabcolsep}{.7 pt}
\centering
\caption{Lowest exponent matrices of codes with $g=12$ and $c=4$}
\label{Table:TTab8}
\vspace{-.7em}
\begingroup\fontsize{6.5pt}{7pt}
\begin{tabular}{|@{}c@{}|@{}c@{}|@{}l@{}|}
\hline
  $a$ & $m_{h}$ &~~~~~~~Exponent Matrix   \\
  \hline
  $5$ & $\renewcommand{\arraystretch}{0.3}\begin{array}{@{}c@{}}
245\\
(1903 \text{\cite{sullivan1}} ) \end{array}$  & $\renewcommand{\arraystretch}{0.3}\begin{array}{l}
0,0,28,27,100\\
45,88,245,35,127\\
100,68,0,211,44\\
0,206,245,245,42 
  \end{array}$ \\
\hline
  $6$ & $\renewcommand{\arraystretch}{0.3}\begin{array}{@{}c@{}}
612\\
(2251\text{\cite{sullivan1}} )\end{array}$ & $\renewcommand{\arraystretch}{0.3}\begin{array}{l}
0,0,71,0,0,0\\
0,62,319,192,316,568\\
39,612,0,380,325,97\\
0,612,117,578,601,609
  \end{array}$ \\
\hline
$7$ & $\renewcommand{\arraystretch}{0.3}\begin{array}{@{}c@{}}
 1333\\
(4019 \text{\cite{sullivan1}} )\end{array}$ & $\renewcommand{\arraystretch}{0.3}\begin{array}{l}
0,0,1263,973,0,0,153\\
0,1,1267,982,655,872,15\\
0,465,752,416,1087,39,0\\
1277,642,0,1277,277,0,1333
\end{array}$ \\
\hline
\end{tabular}
\endgroup
\vspace{-.7em}
\end{table} 

\subsection{Latency and complexity performance}

We denote the smallest lifting degree and syndrome former memory order found through the considered approach as $\tilde{N}$ and $\tilde{m}_h$, respectively.
The corresponding minimum values obtained through previous approaches are instead denoted as $N^*$ and $m_h^*$, respectively.
According to \eqref{eq:lambdaBP}, we can compute the ratio of the decoding latency of the newly designed QC-LDPC block codes over that of previous QC-LDPC block codes as
\begin{equation}
\Theta_N=\frac{\tilde{N}}{N^*}
\label{eq:ratioQC}.
\end{equation}
Similarly, starting from \eqref{cas}, we can compute the ratio of the decoding latency and per-output-bit complexity achieved by the newly designed SC-LDPC-CCs over classical SC-LDPC-CCs as
\begin{equation}
\Theta_{m_h}=\frac{\tilde{m}_h+1}{m_h^*+1}.
\label{eq:ratios}
\end{equation}
The smaller the values of $\Theta_N$ and $\Theta_{m_h}$, the larger the improvement over classical codes.
The smallest values of $\Theta_N$ and $\Theta_{m_h}$ we have obtained in our examples are $0.47$ and $0.23$, respectively.

\subsection{Error rate performance}

\begin{figure}
\centering
\includegraphics[width=78mm,keepaspectratio]{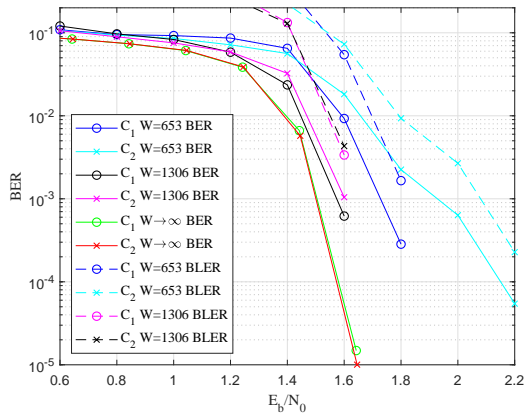}
\vspace{-.7em}
\caption{Simulated performance of SC-LDPC-CCs with $g=12$ as a function of the signal-to-noise ratio.}
\label{fig:perf}
\vspace{-.4em}
\end{figure}

In this section we assess the performance of the new codes in terms of bit error rate (BER) and block error rate (BLER) through Monte Carlo simulations of binary phase shift keying modulated transmissions.
We consider an SC-LDPC-CC designed according to \cite{Tadayon2018} (noted as $C_{1}$) under full-size BP decoding and SW decoding with different window sizes. Notice that the BLER refers to the block of $a$ target symbols decoded within each decoding window.
The maximum number of decoding iterations is $100$, and the actual number of iterations is equal to the maximum for SW decoding (which does not use any stopping criteria based on parity-checks).
We also consider an SC-LDPC-CC (noted as $C_{2}$) with the same code rate and girth as $C_{1}$, obtained by unwrapping a QC-LDPC block code designed following \cite{Bocharova1}. The parameters of the two codes and the value of $\Theta_{m_h}$ are shown in Table \ref{table:Tabparamsc}. Their performance is shown in Figure \ref{fig:perf}. 
We notice that $C_{\mathrm{1}}$ and $C_{\mathrm{2}}$ have almost coincident performance when $W\rightarrow \infty$. However, when the window size is relatively small, $C_{1}$ outperforms $C_{2}$. This happens because the small window sizes imply $\alpha<5$ for $C_{2}$, thus degrading performance, according to the discussion in Section \ref{subsec:4b}. 
The code $C_{1}$, instead, has a value of $v_s$ which is about twice as small as that of $C_{\mathrm{2}}$, yielding values of $\alpha$ that are about twice as big as those of $C_{2}$, and this results in a better performance under SW decoding.

For completeness, let us consider the codes in Table \ref{Table:TTab7} with rate $R=\frac{a-3}{a}$, for $a=6,7,8$, denoted as $\bar{C}_a$, and assess the performance loss $\Delta_{\mathrm{dB}}$ in which they incur at $\mathrm{BER}=10^{-4}$, under full-size BP decoding, with respect to the most compact codes in the literature, noted as $C^*$.
The results are reported in Table \ref{table:Tabloss}, from which we observe that the price paid in terms of $\Delta_{\mathrm{dB}}$ for the corresponding benefit in terms of $\Theta_{m_h}$ is very small.

\begin{table}[!t]
\renewcommand{\arraystretch}{1}
\setlength{\tabcolsep}{1.5 ex}
\caption{Parameters of the considered SC-LDPC-CCs with $R=\frac{5}{8}$}
\vspace{-.7em}
\label{table:Tabparamsc}
\centering
\begin{tabular}{|c|c|c|c|c|c|c|c|c|}
\hline
Code & $a$ & $c$ & $m_h$ & $v_s$ & $g$&$\Theta_{m_h}$\\ \hline\hline
$C_{1}$ & $8$ & $3$ & $297$ & $2384$ & $12$ & \multirow{2}{*}{$0.4563$} \\
\cline{1-6}
$C_{2}$ & $8$ & $3$ & $652$ & $5224$ & $12$& \\ \hline
\end{tabular}
\end{table}

\begin{table}[!t]
\renewcommand{\arraystretch}{1}
\setlength{\tabcolsep}{1.5 ex}
\caption{$\Delta_{\mathrm{dB}}$ and $\Theta_{m_h}$ of SC-LDPC-CCs with $c=3$ and $g=12$}
\vspace{-.7em}
\label{table:Tabloss}
\centering
\begin{tabular}{|c|c|c|c|c|c|c|c|c|c|}
\hline
 & $\bar{C}_6$ & $C^*_{6}$ & $C_7$ & $C^*_{7}$ & $C_8$ & $C^*_{8}$\\ \hline\hline
$\Delta_{\mathrm{dB}}$ & \multicolumn{2}{c|}{$0.03$} & \multicolumn{2}{c|}{$0.005$} & \multicolumn{2}{c|}{0.05}\\
\hline
$\Theta_{m_h}$ &\multicolumn{2}{c|}{$0.81$}&\multicolumn{2}{c|}{$0.75$}&\multicolumn{2}{c|}{$0.67$} \\ \hline
\end{tabular}
\end{table}

\section{Conclusion} \label{Sec6}

Compact QC-LDPC block codes and SC-LDPC-CCs designed through a novel approach based on SMCs allow achieving reduced decoding latency and per-output-bit decoding
complexity, while exhibiting comparable error rate performance
with respect to previous solutions.



%

\end{document}